\documentclass[12pt, draftclsnofoot, onecolumn]{IEEEtran}

\usepackage{graphics} 
\usepackage{epsfig} 

\usepackage{amsmath} 
\usepackage{amssymb}  
\usepackage{amsfonts}
\usepackage{amsthm}
\usepackage[ruled,vlined]{algorithm2e}
\usepackage{graphicx}
\usepackage{epstopdf}
\usepackage{color}

\usepackage{bm}
\usepackage{subfigure}
\usepackage{cases}

\begin{document}

\title{End-to-end Throughput Maximization for Underlay Multi-hop Cognitive Radio Networks with RF Energy Harvesting}

\author{
        Chi~Xu,~
        Meng~Zheng,~
        Wei~Liang,~
        Haibin~Yu,~
         and~Ying-Chang Liang,~

\thanks{C. Xu is with the State Key Laboratory of Robotics and the Key Laboratory of Networked Control Systems, Shenyang Institute of Automation, Chinese Academy of Sciences, Shenyang 110016, China, and also with University of Chinese Academy of Sciences, Beijing 100049, China (e-mail: xuchi@sia.cn).}
\thanks{M. Zheng, W. Liang and H. Yu are with the State Key Laboratory of Robotics and the Key Laboratory of Networked Control Systems, Shenyang Institute of Automation, Chinese Academy of Sciences, Shenyang 110016, China (e-mail: zhengmeng\_6@sia.cn; weiliang@sia.cn; yhb@sia.cn).}
\thanks{Y.-C. Liang is with School of Electrical and Information Engineering, University of Sydney, NSW 2006, Australia, and also with University of Electronic Science and Technology of China, Chengdu 611731, China. (e-mail: liangyc@ieee.org).}
}

\maketitle

\begin{abstract}
This paper studies a green paradigm for the underlay coexistence of primary users (PUs) and secondary users (SUs) in energy harvesting cognitive radio networks (EH-CRNs), wherein battery-free SUs capture both the spectrum and the energy of PUs to enhance spectrum efficiency and green energy utilization.
To lower the transmit powers of SUs, we employ multi-hop transmission with time division multiple access, by which SUs first harvest energy from the RF signals of PUs and then transmit data in the allocated time concurrently with PUs, all in the licensed spectrum.
In this way, the available transmit energy of each SU mainly depends on the harvested energy before the turn to transmit, namely energy causality.
Meanwhile, the transmit powers of SUs must be strictly controlled to protect PUs from harmful interference.
Thus, subject to the energy causality constraint and the interference power constraint, we study the end-to-end throughput maximization problem for optimal time and power allocation.
To solve this nonconvex problem, we first equivalently transform it into a convex optimization problem and then propose the joint optimal time and power allocation (JOTPA) algorithm that iteratively solves a series of feasibility problems until convergence.
Extensive simulations evaluate the performance of EH-CRNs with JOTPA in three typical deployment scenarios and validate the superiority of JOTPA by making comparisons with two other resource allocation algorithms.
\end{abstract}


\IEEEpeerreviewmaketitle

\section{Introduction}
\IEEEPARstart{E}{nergy}
harvesting is an appealing technique that will help solve energy-constrained problems of wireless networks since it can virtually provide perpetual energy supply without manual battery recharging or replacement.
In particular, radio-frequency (RF) energy harvesting \cite{Luxiao15} is more flexible and sustainable than conventional solar or wind energy harvesting since more and more wireless transmitters are deployed and the RF signals radiated by ambient transmitters are consistently available.
Although the energy provided by non-intended energy harvesting is still limited, it is enough for low-power networks, such as sensor networks \cite{Sudevalayam11}.
For high-power networks with more energy requirements, intended power beacons (PBs) are designed for wireless energy transfer (WET) \cite{Zhongcaijun15}, \cite{Guojing15} and wireless powered communication networks (WPCNs) \cite{Ju14TWC}--\cite{Wuqingqing16}.
Furthermore, simultaneous wireless information and power transfer (SWIPT) is studied when the information receiver and the energy receiver are separated in different terminals or co-located in the same terminal \cite{Zhangrui13}.

When energy harvesting is adopted in cognitive radio networks (CRNs), namely energy harvesting CRNs (EH-CRNs), the RF signals radiated by the primary users (PUs) are no longer interference for the secondary users (SUs), but can be regarded as a green energy source for energy harvesting. In this way, SUs can utilize both the spectrum and the energy of PUs. The goal of this paper is to study a green coexistence paradigm for PUs and SUs to enhance spectrum efficiency and green energy utilization simultaneously.

To be specific, we consider an underlay EH-CRN with battery-free SUs that perform multi-hop transmission to lower their transmit powers. By time division multiple access (TDMA), each SU on the multi-hop path first harvests energy from the RF signals of primary transmitter (PT) and then transmits in the allocated time sequentially, both in the licensed spectrum.
The harvested energy is transiently kept by a supercapacitor which has the characteristics of small form factor and fast charge-discharge \cite{Sudevalayam11}.
However, due to the leakage of supercapacitor and the absence of energy storage or management, the possible remaining energy or the harvested energy after transmission cannot be used in the next communication cycle. Thus, the available energy for each SU is mainly determined by the energy harvested before transmission, which is termed as energy causality \cite{Kangxin15}.
In this way, the transmit powers of SUs are subject to not only the interference power constraint of primary receiver (PR) as in conventional underlay CRNs, but also the energy causality constraint imposed by energy harvesting.
Typical applications of this EH-CRN can be cognitive radio sensor networks, in which each SU is a cognitive sensor that transmits data via multiple hops to the sink.

With the formulated green coexistence paradigm, we investigate optimal resource allocation for end-to-end throughput maximization. Specifically, by taking into account the energy causality constraint and the interference power constraint, we formulate an end-to-end throughput maximization problem with respect to time and power allocation. However, this is a nonconvex problem. Thus, we first transform it into an equivalent yet convex optimization problem, and then propose the joint optimal time and power allocation (JOTPA) algorithm to solve it. The JOTPA algorithm decomposes the transformed problem into a series of feasibility problems, each with a given end-to-end throughput, and iteratively solves them by the dual decomposition method until the end-to-end throughput achieves the maximum.
To evaluate the performance of the EH-CRN with JOTPA, we design three typical scenarios for the deployments of EH-CRN and compare JOTPA with two other resource allocation algorithms, namely optimal time and equal power allocation (OTEPA) and equal time and optimal power allocation (ETOPA).

The major contributions of this paper can be summarized as follows:
\begin{itemize}
\item[$\bullet$]
First, different from conventional CRNs that take PT as interference or even ignore PT for simplicity, this paper regards PT as a friend that provides both spectrum and energy for SUs.
In doing so, we investigate a novel green coexistence paradigm to enhance spectrum efficiency and green energy utilization. Moreover, this paradigm can be regarded as another type of SWIPT where PR is the information receiver and SUs are the energy receivers.
\item[$\bullet$]
Second, subject to the energy causality constraint and the interference power constraint, we study the end-to-end throughput maximization problem with respect to time and power allocation and propose the JOTPA algorithm to obtain the optimal solution.
To the best of our knowledge, the optimal resource allocation for the end-to-end throughput maximization of underlay multi-hop EN-CRNs has not been studied before.
\item[$\bullet$]
Third, by moving the EH-CRN around PUs, we investigate three typical scenarios to show how the deployments of EH-CRN influence the performance, which can guide us to deploy the EH-CRN properly.
Moreover, to validate the superiority of the proposed algorithm, we compare JOTPA with OTEPA and ETOPA, and demonstrate that JOTPA gains larger end-to-end throughput and higher green energy utilization than OTEPA and ETOPA under all considered scenarios.
\end{itemize}

The rest of this paper is organized as follows.
Section II reviews the related works on EH-CRNs.
Section III presents the green coexistence paradigm of underlay multi-hop EH-CRNs.
Section IV first formulates and transforms the end-to-end throughput maximization problem, and then proposes the JOTPA algorithm to solve it.
Extensive simulation results are presented and analyzed in Section V, and conclusions are drawn in Section VI.
\section{Related Work}
To exploit the spectrum and the energy of PUs, SUs in EH-CRNs can operate in three kinds of paradigms, namely, interweave, overlay and underlay \cite{Goldsmith09}.

In interweave paradigms, SUs first harvest energy and then opportunistically access the licensed spectrum when PUs are detected as inactive.
In \cite{Chung14}, spectrum sensing is optimized to maximize the throughput when SUs harvest energy from ambient energy sources.
Then, a two-dimensional spectrum and power sensing scheme is proposed for the case that both PUs and SUs harvest energy from the same renewable source \cite{Zhangyanyan15}.
Furthermore, a spectrum access scheme is proposed for throughput maximization in \cite{Hoang14}, in which SUs harvest energy from the active PUs.
In \cite{Lee13}, throughput and outage probability are derived for the case that SUs opportunistically harvest energy from the nearby PUs.

In overlay paradigms, SUs consume energy to serve both PUs and SUs provided that there are excellent cooperations between PUs and SUs.
In \cite{Yinsixing14}, \cite{Wangzihao16}, throughput maximization is studied for the case that one SU harvests energy from ambient RF signals, serves as the relay for PUs, and communicates with another SU.
After extending one SU to multiple SUs, relay selection to maximize throughput or sum-throughput is investigated in \cite{Wangying15}, \cite{Kalamkar16}, wherein SUs harvest from PUs or a hybrid access point (H-AP).
In \cite{Lee15}, sum-throughput maximization is studied for overlay EH-CRNs, wherein first H-AP performs WET for multiple SUs as well as information transmission for PUs, and then collects data from SUs.
Similarly, energy efficiency maximization is studied in \cite{Yinsixing17}, wherein uplink scheduling and cooperative power control are considered.

In underlay paradigms, SUs transmit with the harvested energy as long as the interference to PUs does not exceed a tolerable threshold.
In \cite{Lee15}, sum-throughput maximization is also studied for underlay EH-CRNs, wherein both WET and information transmission are performed concurrently with PUs.
For the case that SUs harvest energy from PUs, throughput is optimized under the outage constraint of PUs \cite{Rakovic15} and the interference power constraint of PUs \cite{Zhengmeng16}.
Moreover, outage probability is derived for the case that SUs harvest energy from PUs \cite{Mousavifar14} or other SUs \cite{Yangzheng15}.
Recently, outage minimization is studied for the case that SUs harvest energy from PB \cite{Xuchi16}.
For hybrid interweave and underlay EH-CRNs, the throughput defined in terms of outage probability is optimized in \cite{Usman14}.

By comparing the above paradigms, we can observe that all the harvested energy in underlay EH-CRNs can be utilized for self-sustainability while much of the harvested energy must be spent on the spectrum sensing in interweave EH-CRNs or the cooperation with PUs in overlay EH-CRNs.
Moreover, as underlay EH-CRNs enable the coexistence of PUs and SUs and are mostly applied when PUs always exist, SUs can take PUs as a stable green energy source for sustainable energy harvesting. In this way, SUs can capture the spectrum and the energy of PUs simultaneously. Motivated by the above analysis, we study underlay EH-CRNs to enhance spectrum efficiency and green energy utilization.

Although a few of the aforementioned works have already studied underlay EH-CRNs, they individually focus on different problems.
To summarize, \cite{Lee15}--\cite{Zhengmeng16} investigate the throughput maximization of single-hop EH-CRNs, while \cite{Mousavifar14}, \cite{Yangzheng15} analyze the outage performance of dual-hop EH-CRNs.
In contrast, this paper studies the end-to-end throughput maximization of multi-hop EH-CRNs.
Moreover, different from our recent work \cite{Xuchi16} studying end-to-end outage minimization of multi-hop EH-CRNs with intended WET, this paper studies end-to-end throughput maximization by exploiting green energy.
In addition, although other recent works \cite{Zhongcaijun15}--\cite{Zhangrui13} studying WET, WPCNs or SWIPT can provide on-demand energy by proper antenna or waveform design and power control, spectrum efficiency and green energy utilization are ignored. In contrast, both issues are taken into account in this paper, wherein the EH-CRNs completely lives on PUs whose setup is prescribed without being influenced by the EH-CRNs.

\section{System Model}
\begin{figure}
\begin{center}
\includegraphics[height=8cm]{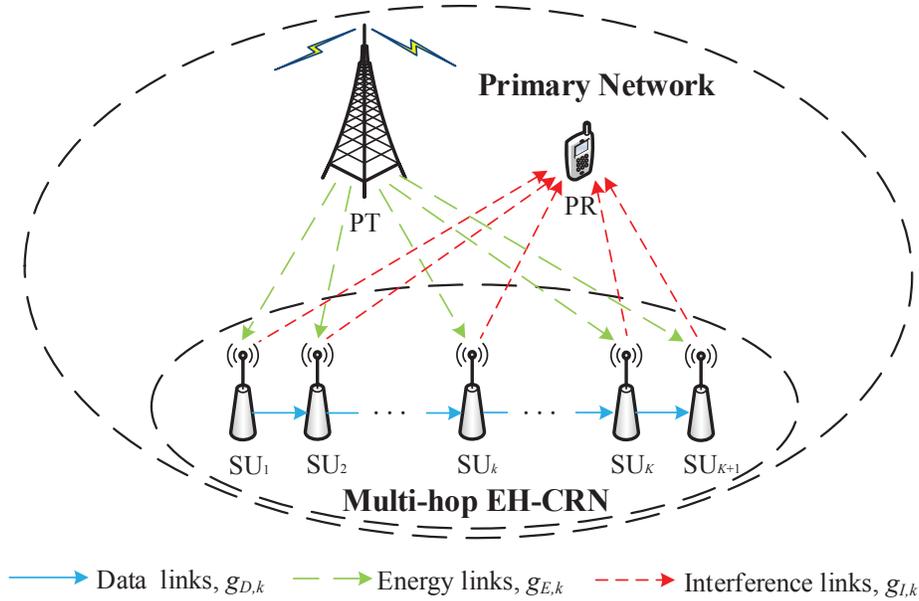}    
\caption{System model.}
\label{fig1}                                 
\end{center}                                 
\end{figure}

As depicted in Fig. \ref{fig1}, an underlay multi-hop EH-CRN with $K+1$ SUs denoted as $SU_k$ ($k=1,...,K+1$), coexists with a pair of PUs, namely PT and PR.
PT is always active in the licensed spectrum to serve PR, and shares the spectrum with SUs under the prescribed interference power constraint \cite{Goldsmith09}, \cite{Lee15}.
All SUs are battery-free without energy storage or management. Meanwhile, SUs do not equipped with any constant energy supplies, but can harvest energy from the RF signals of PT.
Consequently, SUs live on PUs, indicating that the EH-CRN is completely powered by green energy.
However, as non-intended energy harvesting provides limited energy, we employ multi-hop transmission to lower the transmit powers of SUs.
To support the multi-hop transmission, TDMA with perfect time synchronization \cite{Sgora15} is adopted
for the transmission from the source $SU_1$ to the destination $SU_{K+1}$, wherein decode-and-forward (DF) is adopted for relaying.
Furthermore, only one data flow is considered on the multi-hop path to ensure that each SU is allocated with enough energy harvesting time to support its transmission.

\begin{figure}
\begin{center}
\includegraphics[height=6cm]{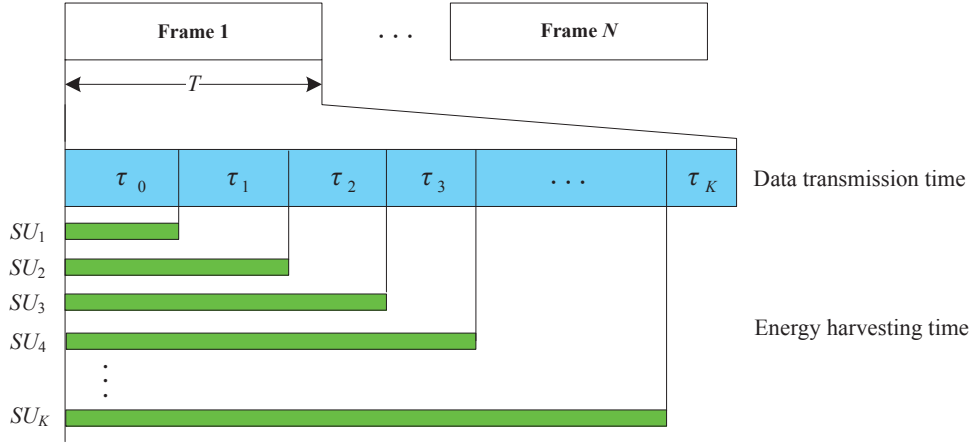}    
\caption{Frame structure.}
\label{fig2}                                 
\end{center}                                 
\end{figure}

With TDMA, a frame with duration $T$ is slotted and allocated to SUs for the $K$-hop transmission as shown in Fig. \ref{fig2}.
The data transmission time allocated to $SU_k$ ($k=1,...,K$) is denoted as $\tau_k$, where $0<\tau_k<T$.
Thus, the total allocated time must satisfy
\begin{equation}\label{equ(1)}
\begin{aligned}
\sum_{k=0}^K \tau_k \leq T,
\end{aligned}
\end{equation}
where $\tau_0$ is designed for the energy harvesting of the source $SU_1$.
It is obvious that the multi-hop transmission can be reduced to a single-hop transmission.

Each SU with one single antenna operates in half-duplex such that it can only transmit or receive at one time.
In this way, each SU works in the harvest-then-transmit fashion. Specifically, $SU_k$ continuously harvests energy from PT before its turn to transmit, wherein the harvested energy is transiently kept in a supercapacitor.
Supercapacitor has the advantages of small form factor and fast charging with theoretically infinite recharge cycles, since it can be trickle-charged by RF signals without complex charging circuitry and has high charge-discharge efficiency without suffering from memory effect \cite{Sudevalayam11}, \cite{Kangxin15}. However, supercapacitor also suffers from high self-discharge.
Thus, due to the leakage of supercapacitor and the absence of energy storage or management, it is reasonable to assume that the possible remaining energy or the energy harvested after transmission in one frame cannot be utilized in the next frame.
In this way, the available energy for each SU in one frame completely depends on the harvested energy before the turn to transmit, namely energy causality.
The energy harvesting time for $SU_k$ is calculated as $\sum_{i=0}^{k-1} \tau_i$.
Obviously, the latter SUs on the multi-hop path have more time for energy harvesting than the former SUs.

The harvested energy of $SU_k$, denoted as $E_k$, is given by
\begin{equation}\label{equ(3)}
\begin{aligned}
E_k=\xi_k P_t g_{E,k} \sum_{i=0}^{k-1} \tau_i, k=1,...,K+1,
\end{aligned}
\end{equation}
where $P_t$ is the transmit power of PT, $g_{E,k}$ is the channel power gain for the energy link from PT to $SU_k$. $0 \leq \xi_k \leq 1$ is the energy harvesting efficiency decided by the design of energy harvester, and $\xi_1=\cdots=\xi_{K+1}=\xi$ is assumed for convenience.
Note that (\ref{equ(3)}) ignores the noise energy as it is too small to contribute to the overall harvested energy.

With the harvested energy, $SU_k$ transmits data to $SU_{k+1}$ in the allocated transmission time $\tau_k$ concurrently with PT in the licensed spectrum.
The received signal $y_{k+1}$ at $SU_{k+1}$ can be expressed as
\begin{equation}\label{equ(2)}
\begin{aligned}
y_{k+1}&=\sqrt{P_t g_{E,k+1}} x_p + \sqrt{P_k g_{D,k}} x_{k} + n_{k+1},
\end{aligned}
\end{equation}
where $n_{k+1} \sim \mathcal{N} (0, \sigma_{k+1}^2)$ is the additive white Gaussian noise (AWGN) at $SU_{k+1}$. Without loss of generality, we assume the noise powers at all SUs are the same, i.e., $\sigma_{1}^2=...=\sigma_{K+1}^2=\sigma^2$.
$x_p$ and $x_k$ are the signals of PT and $SU_k$, wherein $x_p$ fulfilling $\mathbb{E}[|x_p|^2]=1$ and $x_k$ fulfilling $\mathbb{E}[|x_k|^2]=1$ are uncorrelated signals. Thus, $x_p$ is converted into energy by the RF-DC circuit while $x_k$ is decoded and forwarded by $SU_{k+1}$.
$g_{D,k}$ is the channel power gain for the data link from $SU_k$ to $SU_{k+1}$.
$P_k$ is the transmit power of $SU_k$ and must be strictly controlled such that
\begin{equation}\label{equ(4)}
\begin{aligned}
P_k g_{I,k} \leq I_p, k =1,...,K,
\end{aligned}
\end{equation}
where $I_p$ is the peak interference power that PR can tolerate, and $g_{I,k}$ is the channel power gain for the interference link from $SU_k$ to PR.

By taking into account the effects of both large scale path loss and small scale channel fading, we have the channel power gain coefficients as $g_{X,Y}=|h_{X,Y}|^2 (\frac{d_{X,Y}}{d_0})^{-\alpha}$ ($X=E, I, D; Y=k$), where $h_{X,Y}$ and $d_{X,Y}$ are the corresponding channel gain and distance, $d_0$ is the reference distance and $\alpha$ is the path loss exponent.
The channels follow quasi-static block fading (i.e., the channels remain constant during each block transmission but may change from one block to another), and one block fading coincides with a single frame.
We assume that SUs can perfectly evaluate the channel state information (CSI) by channel training and estimation, pilot sensing, direct feedbacks from PUs and SUs, or even indirect feedbacks from a band manager at the beginning of each block \cite{Rakovic15}, \cite{Xuchi16}.

\section{End-to-end Throughput Maximization}
In this section, we investigate the end-to-end throughput maximization problem for optimal resource allocation in the underlay multi-hop EH-CRN.
\subsection{Problem Formulation}
With the harvested energy $E_k$ and the allocated transmission time $\tau_k$,
the achievable throughput for the $k$-th hop transmission from $SU_k$ to $SU_{k+1}$ is calculated as
\begin{equation}\label{equ(5)}
\begin{aligned}
R_k(\tau_k, P_k)=\tau_k \log_2 \left(1+ \frac{P_k g_{D,k}}{\sigma^2}\right), k=1,...,K,
\end{aligned}
\end{equation}
where $\frac{P_k g_{D,k}}{\sigma^2}$ is the signal-to-noise ratio (SNR) at $SU_{k+1}$.
Note that the signal of PT is no longer regarded as interference but is converted into energy.

As the throughput of multi-hop transmission is dominated by the bottleneck link, the end-to-end throughput of the underlay multi-hop EH-CRN is given by
\begin{equation}\label{equ(6)}
\begin{aligned}
R(\bm{\mathcal{\tau}},\bm{\mathcal{P}})=\min_{1 \leq k \leq K} \ R_k(\tau_k, P_k),
\end{aligned}
\end{equation}
where $\bm{\mathcal{\tau}}=[\tau_0,\tau_1,...,\tau_K]$ and $\bm{\mathcal{P}}=[P_1,...,P_K]$ are the vectors of time allocation and power allocation, respectively.

As the end-to-end throughput $R(\bm{\mathcal{\tau}},\bm{\mathcal{P}})$ is a function of both the allocated time $\bm{\mathcal{\tau}}$ and the allocated power $\bm{\mathcal{P}}$, we formulate the following optimization problem with respect to $\bm{\mathcal{\tau}}$ and $\bm{\mathcal{P}}$ to maximize $R(\bm{\mathcal{\tau}},\bm{\mathcal{P}})$:
\begin{equation}\label{equ(7)}
\begin{split}
\max_{\bm{\mathcal{\tau}},\bm{\mathcal{P}}} \ & \min_{1 \leq k \leq K} \ R_k(\tau_k, P_k)
=\tau_k \log_2 \left(1+ \frac{P_k g_{D,k}}{\sigma^2}\right),
\\\text{s.t.} \
& C1': P_k \tau_k \leq \xi P_t  g_{E,k} \sum_{i=0}^{k-1} \tau_i, k=1,...,K,
\\& C2':  P_k g_{I,k} \leq I_p, k=1,...,K,
\\& C3': \sum_{k=0}^{K} \tau_k \leq T,
\\& C4': 0<\tau_k<T, k=0,...,K.
\end{split}
\end{equation}

In problem (\ref{equ(7)}), $C1'$ indicates that, with the allocated time and power, the allocated energy for each SU cannot exceed the harvested energy in a frame; $C2'$ indicates that the peak interference power at PR cannot exceed the tolerable threshold $I_p$.
That is to say, the transmit power of each SU is constrained by the energy causality constraint and the interference power constraint. Moreover, $C3'$ is the total allocated time constraint as (\ref{equ(1)}), while $C4'$ is the allocated time constraint for each SU.

\subsection{Problem Transformation}
Problem (\ref{equ(7)}) is a max-min optimization problem, which maximizes the minimum throughput on the multi-hop path. However, problem (\ref{equ(7)}) is nonconvex since there is the product of optimization variables $\tau_k$ and $P_k$ in $C1'$. To make problem (\ref{equ(7)}) tractable, we introduce new optimization variables $e_k=P_k \tau_k$ ($k=1,...,K$) and transform the problem into the following problem with respect to $\bm{\mathcal{\tau}}$ and $\bm{e}$, where $\bm{e}=[e_1,...,e_K]$ is the vector of the allocated energy for the transmission of each hop.
\begin{equation}\label{equ(8)}
\begin{split}
\max_{\bm{\mathbf{\tau}}, \bm{e}} \ & \min_{1 \leq k \leq K}  \ R_k(\tau_k,e_k)
=\tau_k \log_2 \left(1+ \frac{e_k }{\tau_k} \eta_k \right),
\\\text{s.t.} \
& C1: e_k \leq \xi P_t g_{E,k} \sum_{i=0}^{k-1} \tau_i, k=1,...,K,
\\& C2: e_k g_{I,k} \leq I_p \tau_k, k=1,...,K,
\\& C3: \sum_{k=0}^{K} \tau_k \leq T,
\\& C4: 0<\tau_k<T, k=0,...,K,
\end{split}
\end{equation}
where $\eta_k=\frac{ g_{D,k}}{\sigma^2}$ is defined for convenience, $C3$ and $C4$ are equivalent to $C3'$ and $C4'$ in problem (\ref{equ(7)}), respectively.

\textbf{\emph{Proposition 1:}}
The end-to-end throughput $R(\bm{\mathcal{\tau}}, \bm{e})=\min\limits_{1 \leq k \leq K} R_k(\tau_k,e_k)$ is a jointly concave function of $\bm{\mathcal{\tau}}$ and $\bm{e}$.

\textbf{\emph{Proof:}}
Please refer to Appendix A.
\quad \quad \quad \quad \quad \quad \quad \quad \quad  $\blacksquare$

According to Proposition 1, the objective function in problem (\ref{equ(7)}) with respect to $(\bm{\mathcal{\tau}}, \bm{\mathcal{P}})$ is reformulated as a concave function of $(\bm{\mathcal{\tau}}, \bm{e})$.
Meanwhile, with the help of $\bm{e}$, the energy causality constraint $C1'$ is also converted into an affine constraint as $C1$.
Thus, problem (\ref{equ(8)}) is a convex optimization problem that can be solved by the convex optimization techniques \cite{Boyd04}.
In this way, the original nonconvex optimization problem (\ref{equ(7)}) with respect to time and power allocation is equivalently converted into a convex optimization problem (\ref{equ(8)}) with respect to time and energy allocation.

Based on problem (\ref{equ(8)}), we further introduce a new optimization variable $R$ to reformulate it as the following convex optimization problem.
\begin{equation}\label{equ(9)}
\begin{split}
\max_{\bm{\mathbf{\tau}}, \bm{e}, R} \ & R,
\\\text{s.t.} \
& C0: R_k(\tau_k,e_k) \geq R, k=1,...,K,
\\& C1, C2, C3 \ \text{and} \ C4.
\end{split}
\end{equation}

Problem (\ref{equ(9)}) indicates that, to maximize the end-to-end throughput, we should find the optimal time and energy allocation such that the throughput of each hop is no smaller than $R$.

\subsection{Joint Optimal Time and Power Allocation Algorithm}
To address problem (\ref{equ(9)}), we first provide the following proposition which establishes the relationship between the optimal resource allocation and the maximum end-to-end throughput.

\textbf{\emph{Proposition 2:}}
The maximum end-to-end throughput $R^{*}$ is achieved only when the total frame time is allocated to all SUs on the multi-hop path and each SU obtains the same throughput, namely $\sum_{k=0}^{K} \tau_k^{*} = T$ as well as $R_1(\tau_1^{*},e_1^{*})=...=R_K(\tau_K^{*},e_K^{*})=R^{*}$, where $[\tau_0^{*}, \tau_1^{*},...,\tau_K^{*}]$ and $[e_1^{*},...,e_K^{*}]$ are the optimal time allocation $\bm{\mathcal{\tau}}^{*}$ and the optimal energy allocation $\bm{e}^{*}$, respectively.

\textbf{\emph{Proof:}}
Please refer to Appendix B.
\quad \quad \quad \quad \quad \quad \quad \quad \quad  $\blacksquare$

As shown in Appendix B, for any given total time constraint, we can always find some feasible $R$'s fulfilling $R_1(\tau_1, e_1)=...=R_K(\tau_K, e_K)=R$ by adjusting $\bm{\mathcal{\tau}}$ and $\bm{e}$.
As $R^{*}$ is the maximum of all feasible $R$'s, we first solve the feasibility problem for a given $R$ and then update $R$ to iteratively solve the feasibility problem until $R$ achieves the maximum.

Given $R>0$, the feasibility problem with respect to $\bm{\mathcal{\tau}}$ and $\bm{e}$ is formulated as
\begin{equation}\label{equ(10)}
\begin{split}
\text{find}  \ & (\bm{\mathcal{\tau}},\bm{e}),
\\\text{s.t.} \ & C0, C1, C2, C3 \ \text{and} \ C4.
\end{split}
\end{equation}

Based on the convexity of problem (\ref{equ(9)}), problem (\ref{equ(10)}) is also a convex optimization problem and satisfies Slater's condition \cite{Boyd04}. Thus, the duality gap between the primal problem (\ref{equ(10)}) and its dual problem must be zero, which motivates us to solve the dual problem instead.

The partial Lagrangian function of problem (\ref{equ(10)}) with respect to $C0$ is expressed as
\begin{equation}\label{equ(11)}
\begin{split}
\mathcal{L}(\bm{\mathcal{\tau}},\bm{e},\bm{\mathcal{\lambda}})=-\sum_{k=1}^{K} \lambda_k (R_k(\tau_k,e_k)-R),
\end{split}
\end{equation}
where $\bm{\mathcal{\lambda}}=[\lambda_1,...,\lambda_K]$ with $\lambda_k$ denoting the nonnegative dual variable associated with $C_0$.

Let $\bm{\mathcal{D}}$ denote the feasible set of ($\bm{\mathcal{\tau}}$, $\bm{e}$) specified by $C1-C4$.
The Lagrange dual function of problem (\ref{equ(10)}) is then expressed as
\begin{equation}\label{equ(12)}
\begin{split}
\mathcal{G}(\bm{\mathcal{\lambda}})=\min_{(\bm{\mathcal{\tau}}, \bm{e}) \in \bm{\mathcal{D}}} \mathcal{L}(\bm{\mathcal{\tau}},\bm{e},\bm{\mathcal{\lambda}}),
\end{split}
\end{equation}
which can be regarded as an indicator for the feasibility of problem (\ref{equ(10)}), as provided by the following proposition.

\textbf{\emph{Proposition 3:}}
For a given $R>0$, problem (\ref{equ(10)}) is feasible if and only if there exists $\bm{\mathcal{\lambda}} \geq 0$ such that $\mathcal{G}(\bm{\mathcal{\lambda}}) \leq 0$.

\textbf{\emph{Proof:}}
Please refer to Appendix C.
\quad \quad \quad \quad \quad \quad \quad \quad \quad  $\blacksquare$

Then, the Lagrange dual problem of problem (\ref{equ(10)}) is formulated as $\max \limits_{\bm{\mathcal{\lambda}} \geq 0} \mathcal{G}(\bm{\mathcal{\lambda}})$.

Given $\bm{\mathcal{\lambda}} \geq 0$, $\mathcal{G}(\bm{\mathcal{\lambda}})$ can be reformulated as the following convex optimization problem
\begin{equation}\label{equ(13)}
\begin{split}
\max_{\bm{\mathcal{\tau}},\bm{e}} \ & \sum_{k=1}^{K} \lambda_k R_k(\tau_k,e_k),
\\\text{s.t.} \ & C1, C2, C3 \ \text{and} \ C4,
\end{split}
\end{equation}
where the term $\lambda_k R$ in (\ref{equ(11)}) is ignored since it does not influence the optimality of $(\bm{\mathcal{\tau}},\bm{e})$.

As problem (\ref{equ(13)}) includes both coupling variables in $C2$ and coupling constraints in $C1$ and $C3$, we employ the decomposition method \cite{Palomar06} to solve this problem.
Specifically, we first relax the coupling constraints by dual decomposition and form the partial Lagrangian function of problem (\ref{equ(13)}) as
\begin{equation}\label{equ(14)}
\begin{split}
\mathcal{L}'(\bm{\mathcal{\tau}}, \bm{e}, \bm{\mathcal{\mu}}, \mathcal{\omega})
&=\sum_{k=1}^{K} \lambda_k \tau_k \log_2 \left(1+ \frac{e_k}{\tau_k} \eta_k \right)
-\sum_{k=1}^{K} \mu_k \left(e_k-\xi P_t g_{E,k}\sum_{i=0}^{k-1} \tau_i \right)
-\omega \left( \sum_{k=0}^{K} \tau_k - T \right),
\end{split}
\end{equation}
where $\bm{\mathcal{\mu}}=[\mu_1,...,\mu_K]$, $\mu_k$ denotes the nonnegative dual variable associated with $C1$, and $\omega$ denotes the nonnegative dual variable associated with $C3$.

Then, the Lagrange dual function of problem (\ref{equ(13)}) is given by
$\mathcal{G}'(\bm{\mathcal{\mu}}, \mathcal{\omega})
=\max\limits_{(\bm{\mathcal{\tau}}, \bm{e}) \in \bm{\mathcal{D'}}} \mathcal{L}'(\bm{\mathcal{\tau}}, \bm{e}, \bm{\mathcal{\mu}}, \mathcal{\omega})$, where $\bm{\mathcal{D'}}$ is the set of $(\bm{\mathcal{\tau}}, \bm{e})$ associated with $C2$.
Finally, the Lagrange dual problem of problem (\ref{equ(13)}) is given by $\min\limits_{\bm{\mathcal{\mu}} \geq 0, \mathcal{\omega} \geq 0} \mathcal{G}'(\bm{\mathcal{\mu}}, \mathcal{\omega})$.

The following proposition presents the optimal solution to problem (\ref{equ(13)}).

\textbf{\emph{Proposition 4:}}
For given $\bm{\mathcal{\lambda}} \geq 0$ and $\bm{\mathcal{\mu}} \geq 0$, the optimal time and energy allocation is given by
\begin{numcases}
{\tau_k^*=}
-\frac{e_k^* \eta_k \mathcal{W}(\psi_k)}{\mathcal{W}(\psi_k)+1}, \quad k=1,...,K, \label{equ(15)}
\\
T-\sum\limits_{k=1}^{K} \tau_k^*, \quad k=0,\label{equ(16)}
\end{numcases}
\begin{equation}\label{equ(17)}
e_k^*=\min \left(\left(\frac{\lambda_k \tau_k^*}{\ln 2 \mu_k}-\frac{\tau_k^*}{\eta_k}\right)^+, \frac{I_p \tau_k^*}{g_{I,k}}\right), \quad k=1,...,K,
\end{equation}
where $\mathcal{W}(\cdot)$ denotes the Lambert W function \cite{Corless96}, $\psi_k=-\exp \left(-\frac{\ln2}{\lambda_k} \xi P_t \sum_{j=1}^{k} \mu_j g_{E,j}-1 \right) $ and $(t)^+ \triangleq \max(0, t)$.

\textbf{\emph{Proof:}}
Please refer to Appendix D.
\quad \quad \quad \quad \quad \quad \quad \quad \quad  $\blacksquare$

Through the proof of Proposition 4, we can observe that $\mathcal{G}'(\bm{\mathcal{\mu}}, \mathcal{\omega})$ is finally not related to $\mathcal{\omega}$, namely $\mathcal{G}'(\bm{\mathcal{\mu}}, \mathcal{\omega})=\mathcal{G}'(\bm{\mathcal{\mu}})$. Thus, we can first calculate $\bm{\mathcal{\tau}}^{*}$ and $\bm{e}^{*}$ for a given $\bm{\mathcal{\mu}}$, and then update $\bm{\mathcal{\mu}}$ by sub-gradient algorithms until $\mathcal{G}'(\bm{\mathcal{\mu}})$ achieves the minimum.
Specifically, we iteratively calculate $\tau_k$ and $e_k$ for $k=1,...,K$ using (\ref{equ(15)}) and (\ref{equ(17)}) by fixing one of them at one time and the other next time, until they both converge.  With the obtained $[\tau_1^{*},...,\tau_K^{*}]$, we can calculate $\tau_0^{*}$ by (\ref{equ(16)}) as $\sum_{k=0}^K \tau_k^{*}=T$.
In doing so, we obtain ($\bm{\mathcal{\tau}^{*}}$, $\bm{e}^{*}$) for the given $\bm{\mathcal{\mu}}$.
Then, we calculate the optimal dual variable $\bm{\mathcal{\mu}^{*}}$ that minimizes $\mathcal{G}'(\bm{\mathcal{\mu}})$ by the ellipsoid method, where the sub-gradient of $\mathcal{G}'(\bm{\mathcal{\mu}})$ at $\mu_k$ is calculated as
\begin{equation}\label{equ(18)}
\nabla \mu_k=e_k^{*} - \xi P_t g_{E,k} \sum_{i=0}^{k-1} \tau_i^{*}.
\end{equation}

In this way, we solve problem (\ref{equ(13)}) by Proposition 4 for the given $\bm{\mathcal{\lambda}}$ and obtain ($\bm{\mathcal{\tau}^{*}}$, $\bm{e}^{*}$) with which $R_k(\tau_k^{*},e_k^{*})$ can be calculated.
Then, we employ (\ref{equ(11)}) to calculate $\mathcal{G}(\bm{\mathcal{\lambda}})$ given by (\ref{equ(12)}) and further check the feasibility of ($\bm{\mathcal{\tau}^{*}}$, $\bm{e}^{*}$) for the given $R$ by Proposition 3.
Specifically, if $\mathcal{G}(\bm{\mathcal{\lambda}})>0$, problem (\ref{equ(10)}) is infeasible, which means $R_k(\tau_k^{*}, e_k^{*})<R$. Thus, we should decrease $R$ and solve the feasibility problem (\ref{equ(10)}) again. However, if $\mathcal{G}(\bm{\mathcal{\lambda}}) \leq 0$, problem (\ref{equ(10)}) is feasible by Proposition 3, which means $R_k(\tau_k^{*}, e_k^{*}) \geq R$ for the given $\bm{\mathcal{\lambda}}$. Therefore, we should further update $\bm{\mathcal{\lambda}}$ until it converges to $\bm{\mathcal{\lambda}}^{*}$ that maximizes $\mathcal{G}(\bm{\mathcal{\lambda}})$.
The update method can also be the ellipsoid method, where the sub-gradient of $\mathcal{G}(\bm{\mathcal{\lambda}})$ at $\lambda_k$ is given by
\begin{equation}\label{equ(19)}
\begin{split}
\nabla \lambda_k=\tau_k^{*} \log_2 \left(1+ \frac{e_k^{*}}{\tau_k^{*}} \eta_k \right)-R.
\end{split}
\end{equation}
If $\mathcal{G}(\bm{\mathcal{\lambda}}^{*}) \leq 0$ still holds, we should increase $R$ and solve the feasibility problem (\ref{equ(10)}) again; otherwise, we should decrease $R$ and solve the feasibility problem (\ref{equ(10)}) again.

By this means, $R$ is updated and the feasibility problem (\ref{equ(10)}) is iteratively solved until $R$ achieves the maximum within a prescribed error threshold $\delta$.
The method to update $R$ can be a bisection search.
With the obtained optimal time and energy allocation $(\bm{\mathcal{\tau}}^{*}, \bm{e}^{*})$, we can further obtain the optimal power allocation $\bm{\mathcal{P}^{*}}=[P_1^{*},P_2^{*},...,P_K^{*}]$, where $P_k^{*}=\frac{e_k^{*}}{\tau_k^{*}}$ ($k=1,...,K$).

The above process is summarized in Algorithm \ref{alg}.
This algorithm can converge to the optimum due to the following reasons.
First, as $\bm{\mathcal{\tau}}$ and $\bm{e}$ fulfilling the KKT condition are iteratively optimized, they can converge to the optimum for the given $\bm{\mathcal{\lambda}}$ and $\bm{\mathcal{\mu}}$. Then, due to the convex nature of problems (\ref{equ(10)}) and (\ref{equ(13)}), by utilizing the ellipsoid method to update $\bm{\mathcal{\lambda}}$ and $\bm{\mathcal{\mu}}$, we can guarantee that the solution converges to the optimum for the given $R$. Finally, $R$ will also converge to the optimum as the bisection search method to update $R$ is with guaranteed convergence and optimality. As each loop in Algorithm \ref{alg} guarantees the convergence to the optimum, the obtained solution is optimal.

\begin{algorithm}
\LinesNumbered
\footnotesize
\caption{Joint optimal time and power allocation} \label{alg}
\KwIn{$K$, $T$, $P_t$, $I_p$, $\xi$, $\sigma^2$, $\delta$, $g_{E,k}$, $g_{I,k}$, $g_{D,k}$ for $ k=1,...,K$\;}
\KwOut{($\bm{\mathcal{\tau}^{*}}$, $\bm{\mathcal{P}}^{*}$)\;}
Initialization sufficiently small $R_{\min}$ and sufficiently large $R_{\max}$\;
\Repeat{$R_{\max}-R_{\min} < \delta$}
    {
    Calculate $R=\frac{1}{2}(R_{\min}+R_{\max})$ and initialize $\bm{\mathcal{\lambda}} \geq 0$\;
    \Repeat{$\bm{\mathcal{\lambda}}$ converges to $\bm{\mathcal{\lambda}}^*$}
       {
       Initialize ($\bm{\mathcal{\tau}}$, $\bm{e}$) and $\bm{\mathcal{\mu}} \geq 0$\;
       \Repeat{$\bm{\mathcal{\mu}}$ converges to $\bm{\mathcal{\mu}}^*$}
          {
          \Repeat{($\bm{\mathcal{\tau}}$, $\bm{e}$) converges to ($\bm{\mathcal{\tau}}^*$, $\bm{e}^*$)}
             {
              Calculate $\bm{\mathcal{\tau}}$ by (\ref{equ(15)}) and (\ref{equ(16)})\;
              Calculate $\bm{e}$ by (\ref{equ(17)})\;
             }
          Update $\bm{\mathcal{\mu}}$ by the ellipsoid method with (\ref{equ(18)})\;
          }
        Calculate $\mathcal{G}(\bm{\mathcal{\lambda}})$ by (\ref{equ(12)}) using (\ref{equ(11)})\;
        \If {$\mathcal{G}(\bm{\mathcal{\lambda}}) > 0$}
           {Set $R_{\max} \leftarrow R$; break\;}
        \Else {Update $\bm{\mathcal{\lambda}}$ by the ellipsoid method with (\ref{equ(19)})\;}
    }
    \If {$\mathcal{G}(\bm{\mathcal{\lambda}}^{*}) \leq 0$}
        {Set $R_{\min} \leftarrow R$\;}
    \Else {Set $R_{\max} \leftarrow R$\;}
    }
    Calculate $\bm{\mathcal{P}}^{*}$, where $P_k^{*}=\frac{e_k^{*}}{\tau_k^{*}}$ ($k=1,...,K$)\;
\end{algorithm}

\section{Simulation Results}
In this section, we present simulation results to evaluate the performance of the underlay multi-hop EH-CRN.
We first design three typical scenarios by moving the EH-CRN around PUs to evaluate the impact of different deployments on the performance, and then compare the proposed JOTPA algorithm with OTEPA and ETOPA algorithms to validate the superiority of JOTPA.
Note that all the algorithms must fully consider the impact of time allocation on power allocation, as the harvested energy that can be allocated for transmission completely depends on the time allocation.
Thus, for fair comparison, OTEPA optimizes time allocation by the same method as in Algorithm \ref{alg} and allocates power equally, while ETOPA allocates time equally as in common TDMA networks and optimizes power allocation as in conventional underlay CRNs subject to the peak interference power at PR and the maximum transmit powers of SUs.

\subsection{Simulation Setup}
The simulation parameters and scenarios are set as follows.
Without loss of generality, the length of a frame is normalized to a unit time (i.e. $T=1$), and the energy harvesting efficiency is set to $\xi=0.8$.
The noise power is set to unity as $\sigma^2=1$, while both $P_t$ and $I_p$ are normalized by $\sigma^2$.
The transmit power of PT is set to $P_t=40$ dB.
The path loss exponent is set to $\alpha=2$ \cite{Yangzheng15}.
The fading channels are modeled as independent Rayleigh block fading, as a result of which $|h_{E,k}|^2$, $|h_{I,k}|^2$ and $|h_{D,k}|^2$ are independent exponential random variables.
The reference distance is set to $d_0=1$ m.
The distance between the source $SU_1$ and the destination $SU_{K+1}$ is set to 20 m, while the distance between PT and PR is also set to 20 m.
PT and PR are separately located at (0, 10) and (0, -10) on the y-axis.
We consider a linear multi-hop EH-CRN, wherein relay SUs are equally scattered between $SU_1$ and $SU_{K+1}$.
Then, we move the EH-CRN on the x-axis and consider the following three typical scenarios to comprehensively evaluate the impact of deployments on the performance of EH-CRN.

Scenario 1: $SU_1$ and $SU_{K+1}$ are deployed at (0, 0) and (20, 0), respectively, which corresponds to the case that the former SUs are nearer to PUs than the latter SUs.

Scenario 2: $SU_1$ and $SU_{K+1}$ are deployed at (-10, 0) and (10, 0), respectively, which corresponds to the case that SUs are symmetrically distributed around PUs.

Scenario 3: $SU_1$ and $SU_{K+1}$ are deployed at (-20, 0) and (0, 0), respectively, which corresponds to the case that the latter SUs are nearer to PUs than the former SUs.

\subsection{Comparisons of Different Deployment Scenarios}
\begin{figure}
\begin{center}
\includegraphics[height=6cm]{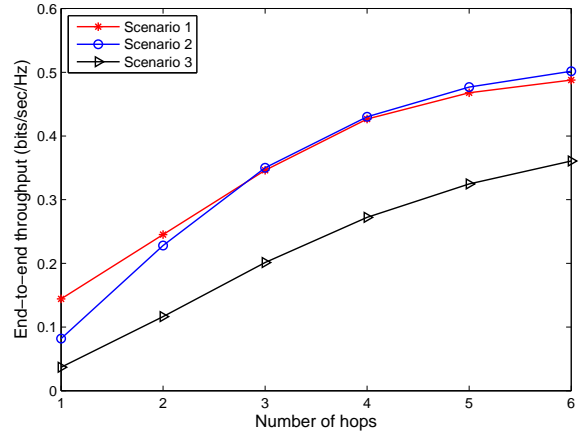}    
\caption{$R^*$ versus $K$ for different scenarios: $P_t=40$ dB, $I_p=10$ dB.}
\label{fig3}                                 
\end{center}                                 
\end{figure}

\begin{figure*}
\subfigure[Energy status in Scenario 1]
{\label{fig4a}\includegraphics[height=4cm]{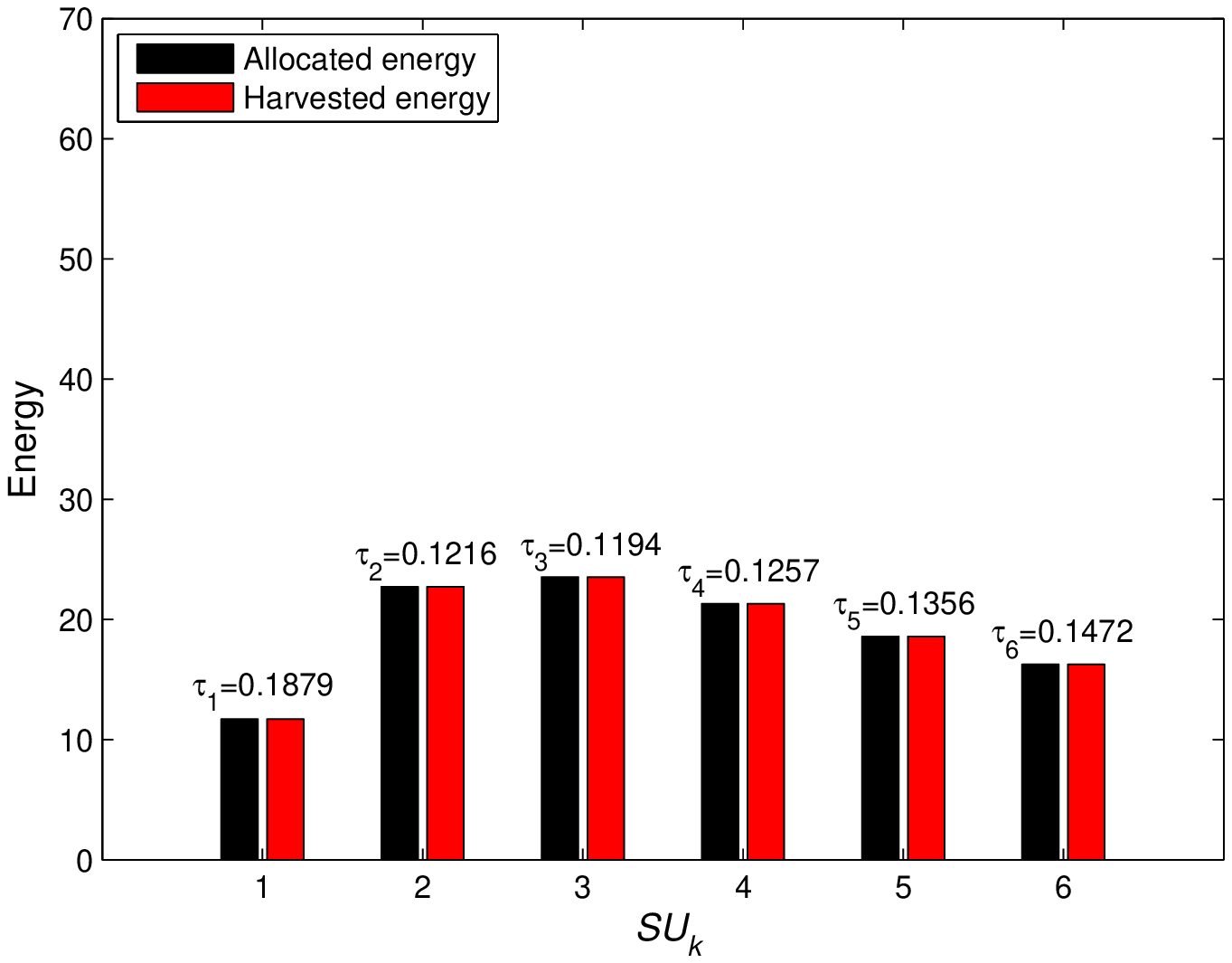}}
\subfigure[Energy status in Scenario 2]
{\label{fig4b}\includegraphics[height=4cm]{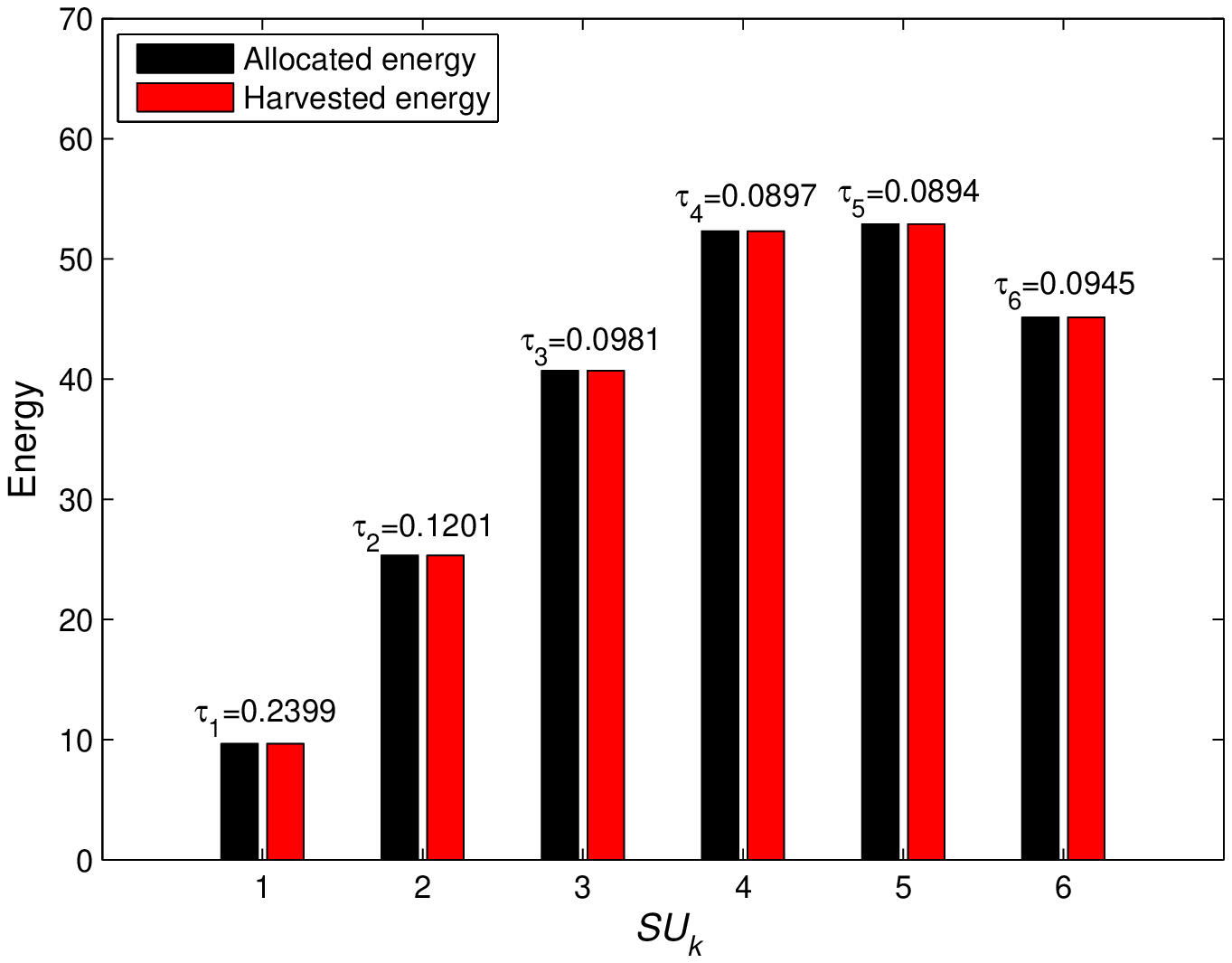}}
\subfigure[Energy status in Scenario 3]
{\label{fig4c}\includegraphics[height=4cm]{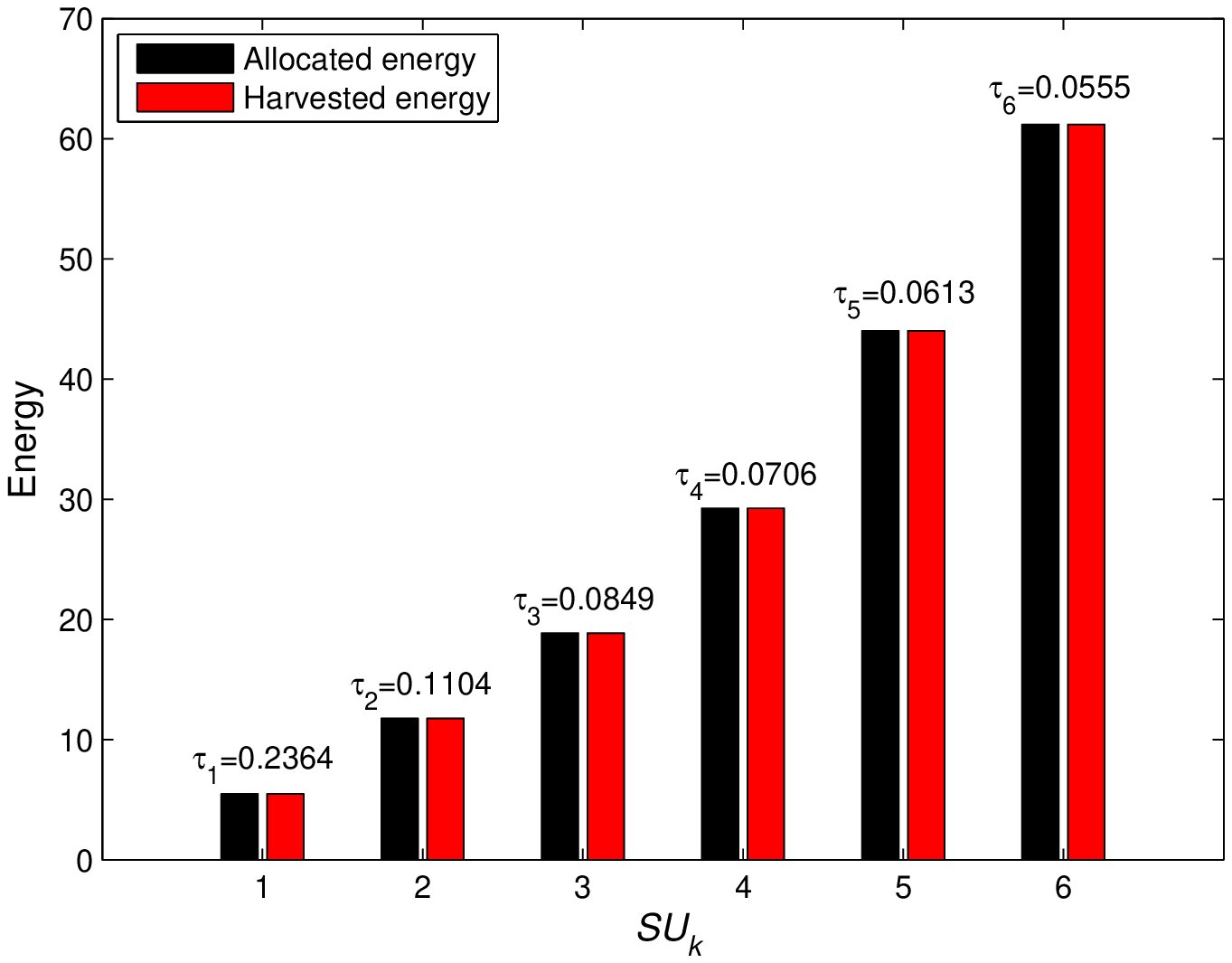}}
\caption{Energy statuses of SUs in different scenarios: $P_t=40$ dB, $I_p=10$ dB.}
\label{fig4}
\end{figure*}

In Fig. \ref{fig3}, we compare the maximum end-to-end throughput $R^*$ of the EH-CRNs with different numbers of hops in the three scenarios.
For the same scenario, $R^*$ increases with the increase of $K$ as more SUs participate in relaying and the path loss decreases with the distance of each hop decreasing.
For the same $K$, it is obvious that $R^*$ of Scenario 1 or 2 is larger than that of Scenario 3.
To explain this phenomenon, we should be aware of the following two facts.
First, in one block fading, the harvested energies of SUs are significantly influenced by the path loss, as a result of which SUs close to PT can harvest more energy.
Second, the harvested energy is monotonically increasing with the increase of energy harvesting time, as a result of which the latter SUs on the multi-hop path can harvest more energy.
In Scenario 1, although the latter SUs suffer from larger path loss than the former SUs, they have more energy harvesting time, which can compensate the impact of path loss.
However, in Scenario 3, the latter SUs suffer from smaller path loss but have more energy harvesting time than the former SUs, which causes the unbalance of the harvested energies among SUs and further the phenomenon that $R^*$ of Scenario 3 is smaller than that of Scenario 1.
Note that Scenario 2 is a compromise of Scenario 1 and Scenario 3, as a result of which $R^*$ of Scenario 2 is also larger than that of Scenario 3.

To help understand the above analysis, we further depict Fig. \ref{fig4} to present the allocated time, the allocated energy and the harvested energy for the six-hop EH-CRN selected from Fig. \ref{fig3}.
Obviously, the energy distribution of Scenario 1 is much more balanced than that of Scenario 3, in which the harvested energy is monotonically increasing. 
Moreover, from Scenarios 1 to 3, the allocated energy and the harvested energy at $SU_1$ are decreasing while those at $SU_6$ are increasing.
These observations are all consistent with the previous analysis for Fig. \ref{fig3}.
In addition, from $SU_1$ to $SU_6$, the allocated transmission time in Scenario 1 is approximate due to the well-balanced energy distribution, while that in Scenario 3 is decreasing even though more energy is harvested.

\begin{figure}
\begin{center}
\includegraphics[height=6cm]{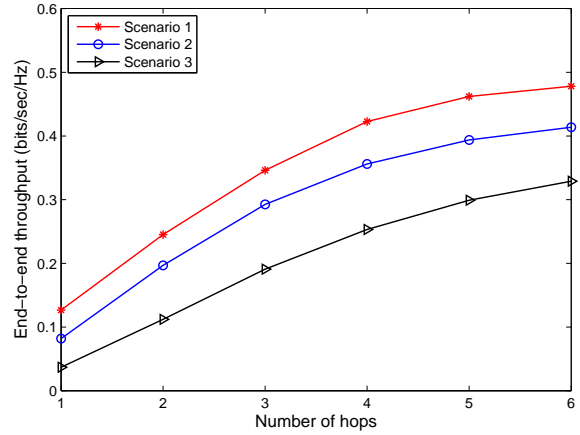}    
\caption{$R^*$ versus $K$ for different scenarios: $P_t=40$ dB, $I_p=0$ dB}.
\label{fig5}                                 
\end{center}                                 
\end{figure}

\begin{figure*}
\subfigure[Energy status in Scenario 1]
{\label{fig6a}\includegraphics[height=4cm]{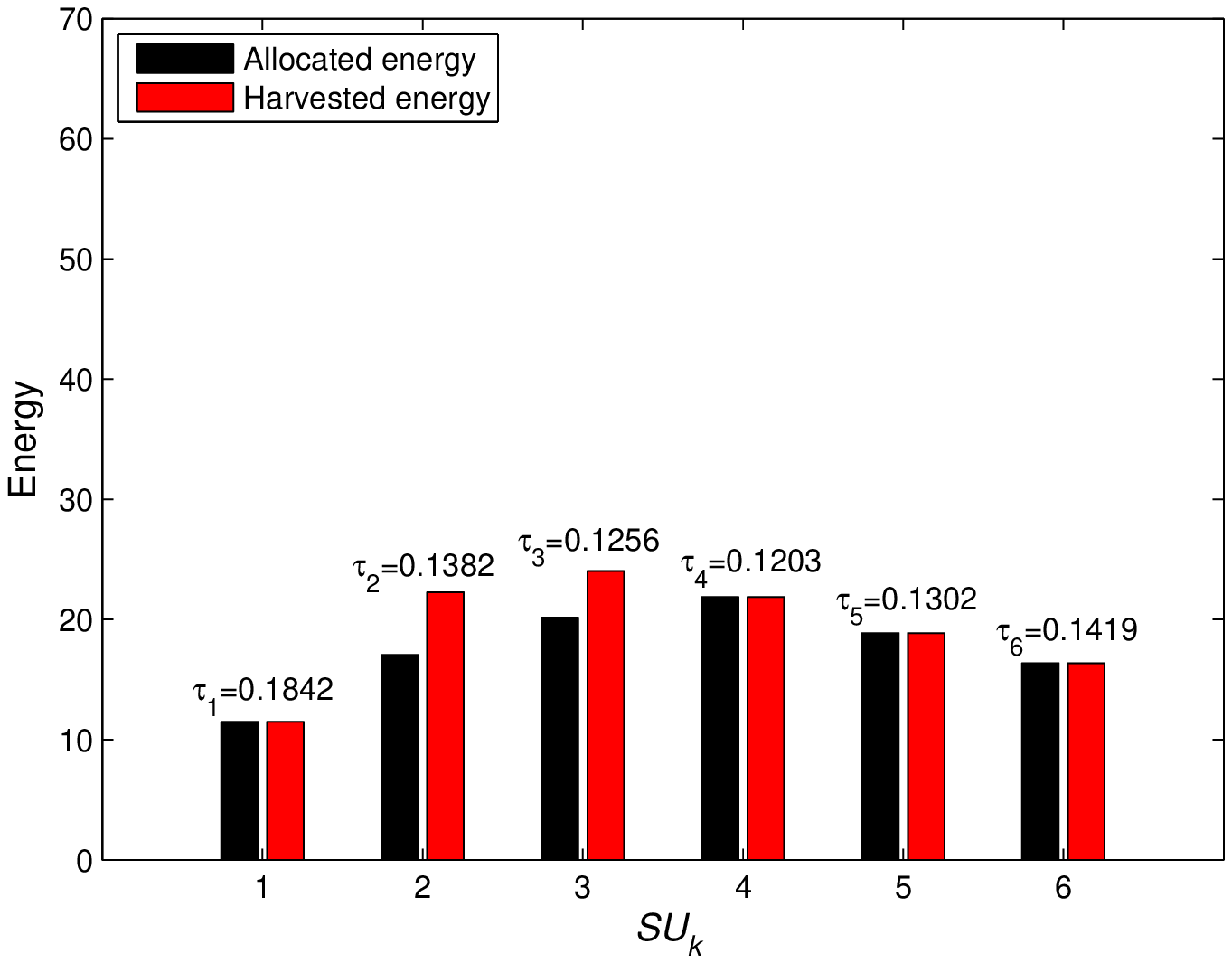}}
\subfigure[Energy status in Scenario 2]
{\label{fig6b}\includegraphics[height=4cm]{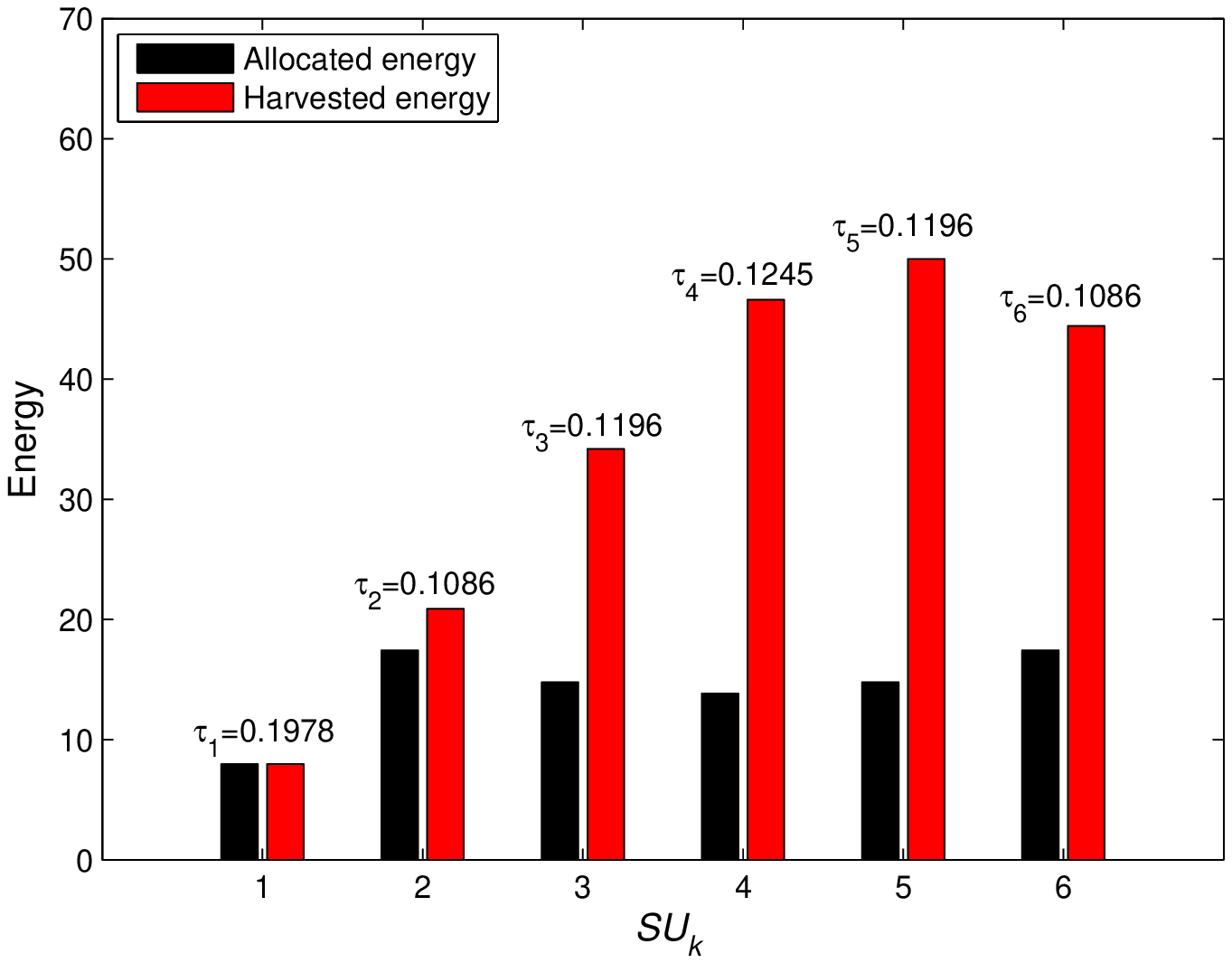}}
\subfigure[Energy status in Scenario 3]
{\label{fig6c}\includegraphics[height=4cm]{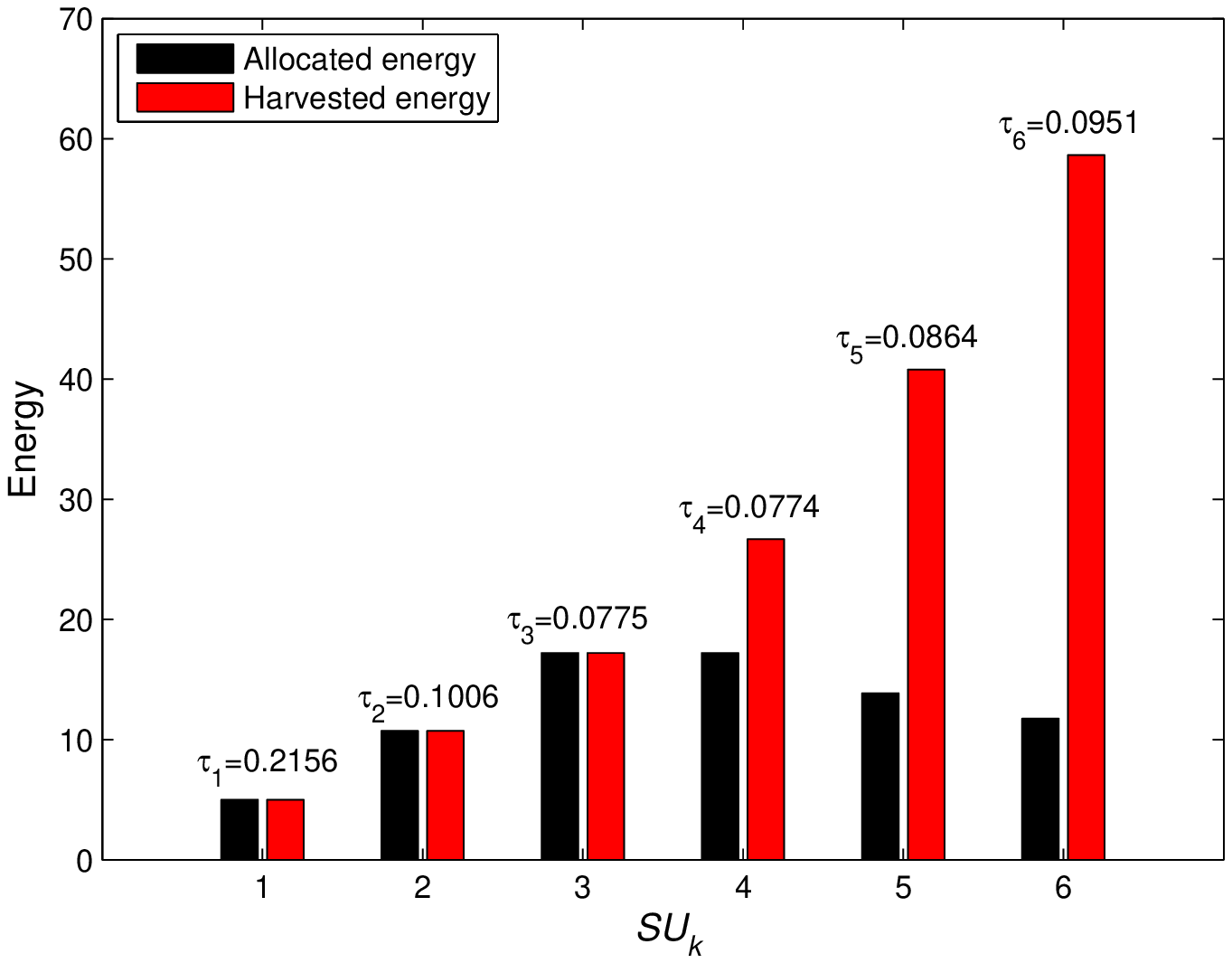}}
\caption{Energy statuses of SUs in different scenarios: $P_t=40$ dB, $I_p=0$ dB.}
\label{fig6}
\end{figure*}
Another significant observation in Fig. \ref{fig4} is that all the harvested energies of SUs are allocated. This is because the interference power constraint of PR is so slack that SUs can consume all the harvested energies to maximize the throughput. 
Therefore, we restrict the interference power constraint of PR to a strict level as $I_p=0$ dB in Fig. \ref{fig5}.
Obviously, the throughput gaps between the three scenarios are enlarged, wherein $R^*$ of Scenario 1 is the largest and $R^*$ of Scenario 3 is the smallest. This is because only a small part of the harvested energies can be allocated for transmission as shown in Fig. \ref{fig6}, while the remaining part will be discharged due to the leakage of supercapacitors.
Thus, we can conclude that when the interference power constraint of PR is strict, the scenario in which former SUs are close to PT can benefit more from the green energy and achieve larger end-to-end throughput than the scenario in which latter SUs are close to PT.

\begin{figure}
\begin{center}
\includegraphics[height=6cm]{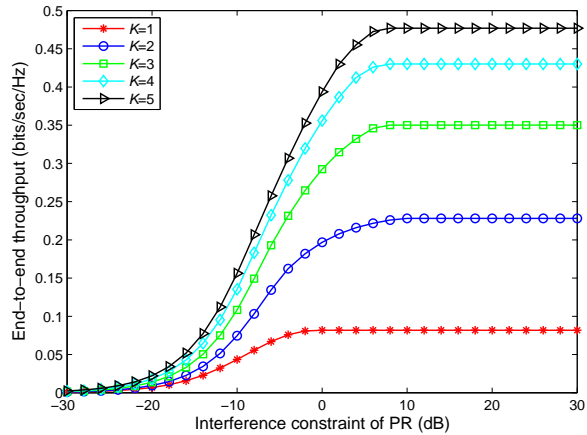}    
\caption{$R^*$ versus $I_p$ for different numbers of hops: $P_t=40$ dB.}
\label{fig7}                                 
\end{center}                                 
\end{figure}

By comparing Figs. \ref{fig3}--\ref{fig6}, we can observe that the interference power constraint of PR significantly influences the end-to-end throughput.
To clearly present the relationship between $R^*$ and $I_p$, we depict Fig. \ref{fig7} for Scenario 2 which employs symmetrical deployment and is selected without loss of generality.
As shown, when $I_p$ is small (e.g., $I_p=-30$ dB), $R^*$ is almost zero since even small transmit powers of SUs will cause intolerable interference to PR. Thus, only a small amount of the harvested energies are utilized for transmission.
Then, with the increase of $I_p$ (i.e., the interference power constraint of PR is slackened), more of the harvested energies can be utilized for transmission and $R^*$ is enhanced correspondingly. However, when $I_p$ is sufficiently large, $R^*$ cannot be further enhanced since the harvested energies by SUs are limited for the given $P_t$.
For this case, the EH-CRN is equivalent to the WPCN without cognitive radio.
For the general case, we set $I_p=5$ dB in the following simulations.

\subsection{Comparisons of Different Resource Allocation Algorithms}
\begin{figure}
\begin{center}
\includegraphics[height=6cm]{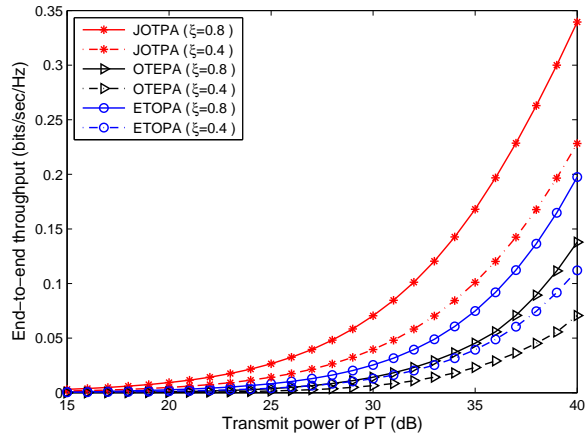}    
\caption{$R^*$ versus $P_t$ for different algorithms: $I_p=5$ dB, $K=3$.}
\label{fig8}                                 
\end{center}                                 
\end{figure}

In Fig. \ref{fig8}, we compare JOTPA with OTEPA and ETOPA for three-hop EH-CRNs with different energy harvesting efficiencies deployed in Scenario 2.
It is obvious that JOTPA gains larger $R^*$ than OTEPA and ETOPA for the same $\xi$.
The reasons are explained as follows.
First, although OTEPA performs optimal time allocation, it does not allocate powers according to CSI and $I_p$. Thus, SUs cannot enhance their transmit powers adaptively even if the harvested energies are sufficient.
Second, as the allocated time by ETOPA is not optimized, the harvested energies for some SUs are not enough to reach the maximum available transmit powers constrained by $I_p$, while for other SUs are so much that cannot be utilized but discharged.
However, no matter which algorithm is employed, $R^*$ always increases with the increase of $P_t$ or $\xi$ since more energies can be harvested and utilized for transmission under the constraint of the same $I_p$.
Moreover, the throughput gaps from OTEPA and ETOPA to JOTPA are enlarging with the increase of $P_t$, which validates the superiority of the proposed JOTPA algorithm.

\begin{figure*}
\subfigure[Energy status of $SU_1$]
{\label{fig9a}
\includegraphics[height=3.8cm]{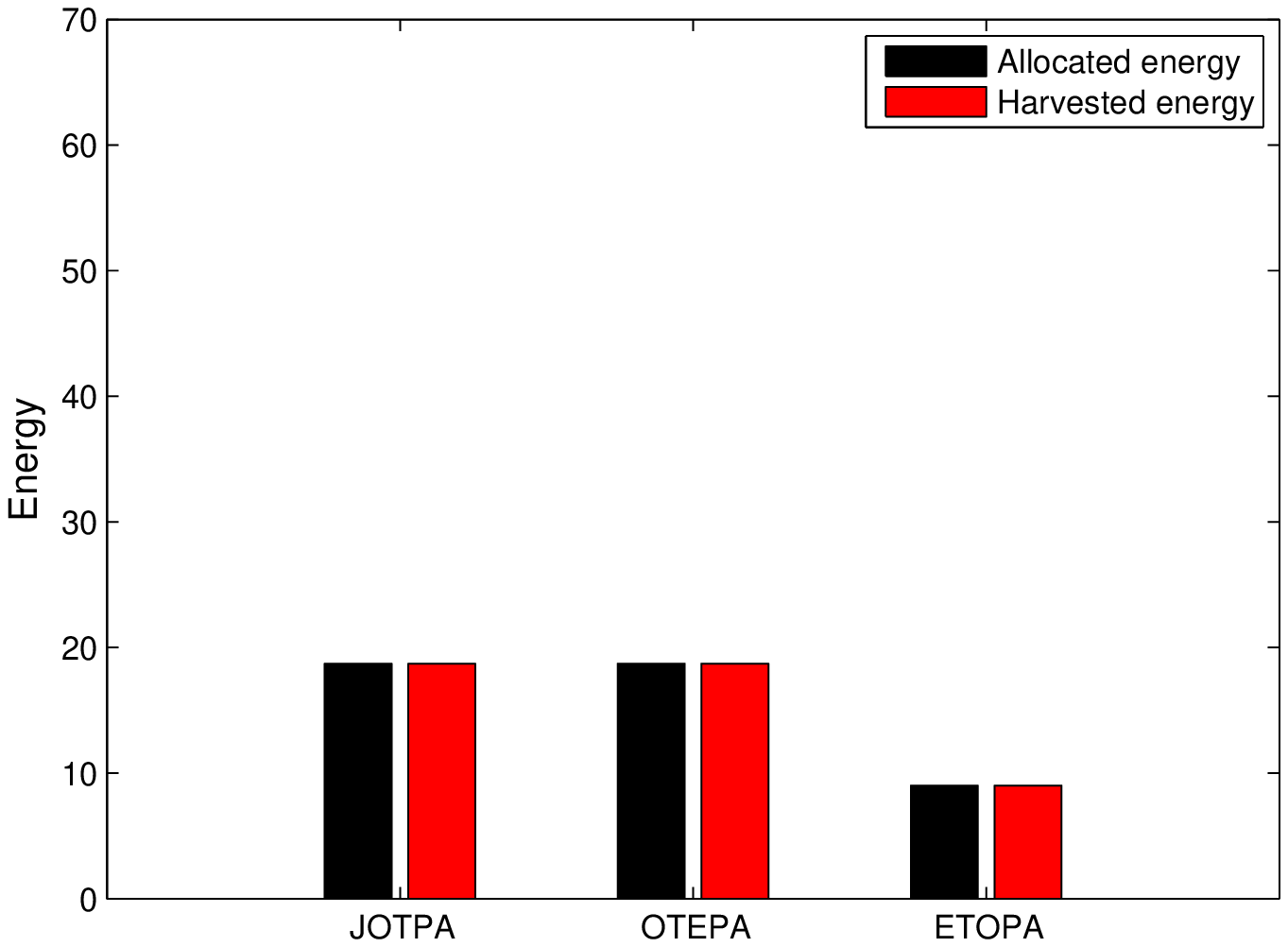}}
\subfigure[Energy status of $SU_2$]
{\label{fig9b}
\includegraphics[height=3.8cm]{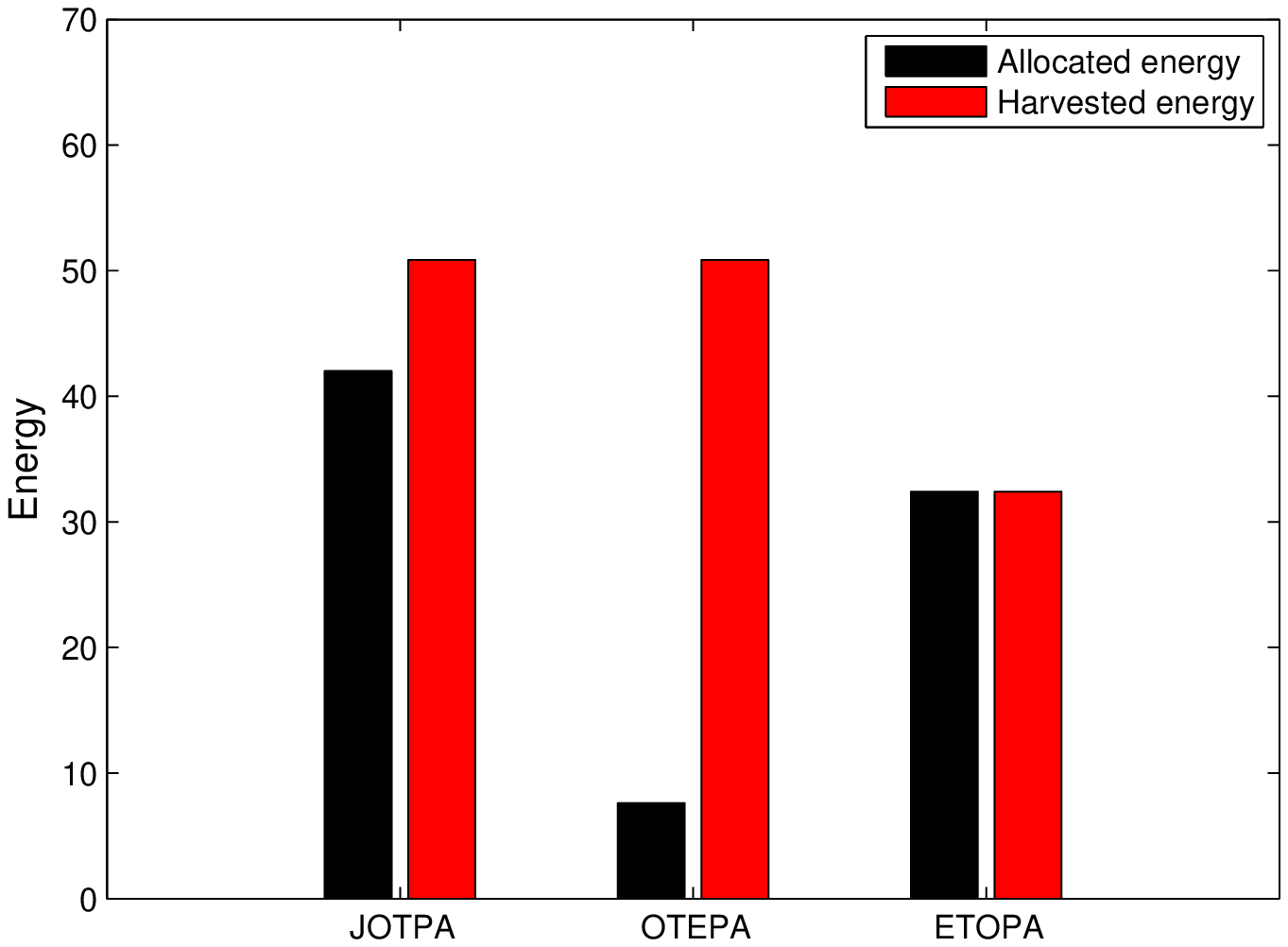}}
\subfigure[Energy status of $SU_3$]
{\label{fig9c}
\includegraphics[height=3.8cm]{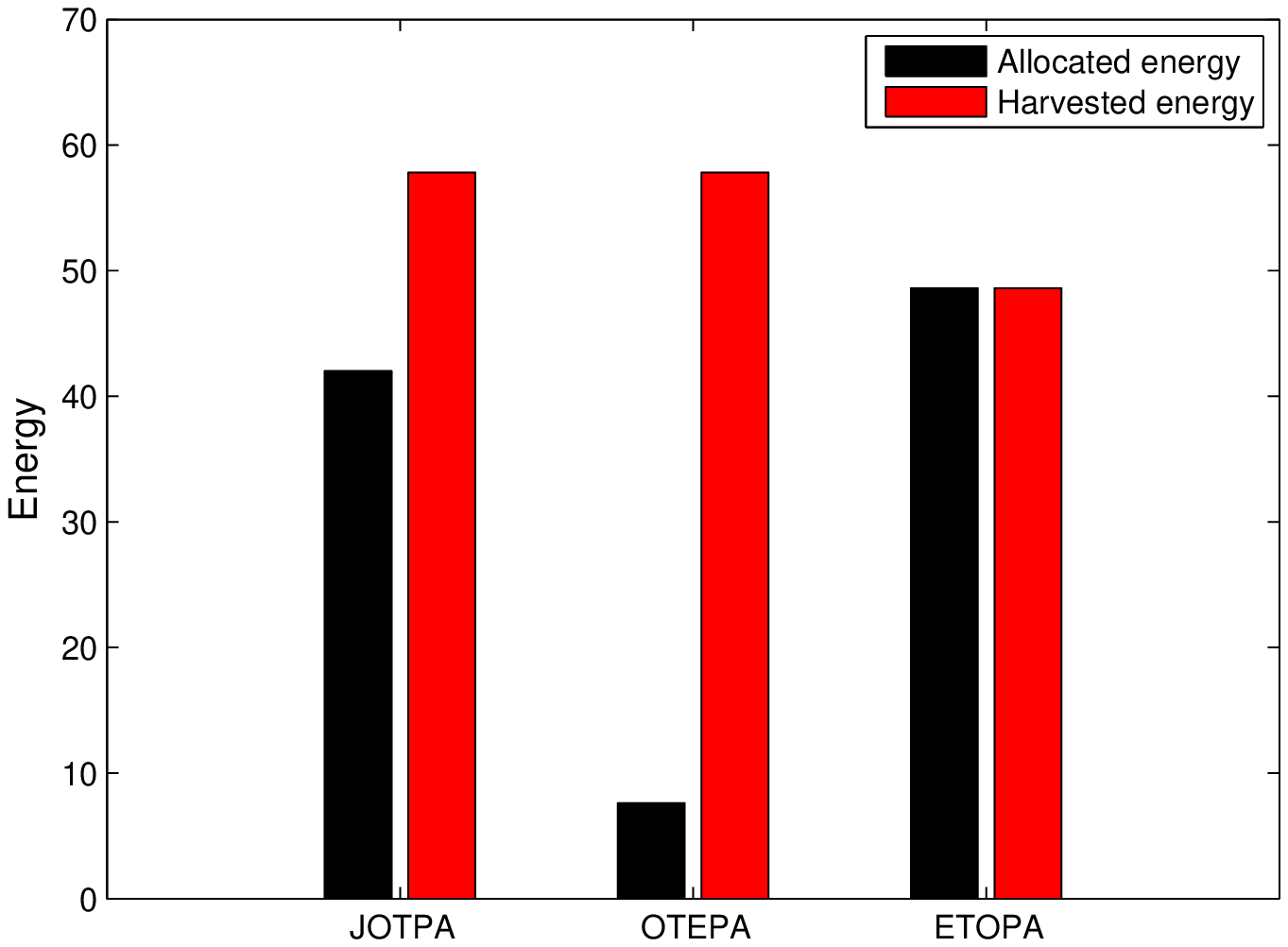}}
\caption{Energy statuses of SUs with different algorithms in Scenario 2: $P_t=40$ dB, $I_p=5$ dB, $K=3$.}
\label{fig9}
\end{figure*}

More specifically, Fig. \ref{fig9} compares the energy statuses of SUs in three-hop EH-CRNs with different resource allocation algorithms.
It is observed that the harvested energies of SUs by ETOPA are much smaller than those by JOTPA and OTEPA, wherein the harvested energies by JOTPA and OTEPA are the same as their time allocations are optimized by the same method.
However, the total allocated energies of all SUs by OTEPA and ETOPA are much smaller than that by JOTPA. Thus, we can conclude that JOTPA outperforms OTEPA and ETOPA in green energy utilization.

\begin{figure}
\begin{center}
\includegraphics[height=6cm]{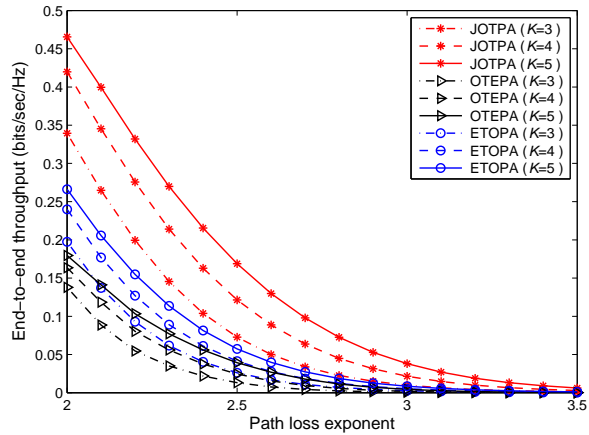}    
\caption{$R^*$ versus $\alpha$ for different algorithms: $P_t=40$ dB, $I_p=5$ dB}.
\label{fig10}                                 
\end{center}                                 
\end{figure}

In Fig. \ref{fig10}, we compare the three algorithms for the EH-CRNs with different numbers of hops.
For any $K$, when $\alpha$ is small, JOTPA always obtains much larger $R^*$ than those by OTEPA and ETOPA. However, when $\alpha$ increases, $R^*$ decreases quickly and the superiority of JOTPA decreases. This is because when the path loss becomes severe, no matter which algorithm is employed, both the energies harvested by transmitters and the information correctly decoded by receivers become scarce.

To overcome the influence of path loss, multi-hop transmission is an efficient method, as clearly shown in Fig. \ref{fig10}, wherein $R^*$ increases with the increase of $K$.
However, $K$ cannot be arbitrarily increased as the throughput gain is not always increasing in $K$. For example, when $\alpha=2$, the throughput gain of JOTPA from $K=3$ to $K=4$ is 23.72\%, while that from $K=4$ to $K=5$ is only 10.86\%. Similar observations can also be obtained from Figs. \ref{fig3}, \ref{fig5} and \ref{fig7}.
The reason is that a frame is divided by $K$ and a large $K$ leads to small $\tau_k$, which decreases $R(\bm{\mathcal{\tau}},\bm{\mathcal{P}})$ as $R(\bm{\mathcal{\tau}},\bm{\mathcal{P}})=\min\limits_{1 \leq k \leq K}(R_k(\tau_k, P_k))$ and $R_k(\tau_k, P_k)$ monotonically increases in $\tau_k$.
We can consider an extreme case when $K \rightarrow \infty$ resulting in $\tau_k \rightarrow 0$, there must be $R(\bm{\mathcal{\tau}},\bm{\mathcal{P}}) \rightarrow 0$.
Thus, too large $K$ cannot benefit more $R^*$, which motivates us to set $K$ properly.
However, the optimal $K$ cannot be calculated by a closed-form expression. Instead, we can obtain the optimal $K$ by a one-dimensional search as $K$ is a finite integer.

\section{Conclusion}
In this paper, we formulated a green coexistence paradigm for underlay multi-hop EH-CRNs, in which battery-free SUs capture both the spectrum and the energy of PUs to enhance spectrum efficiency and green energy utilization.
With this paradigm, we investigated the end-to-end throughput maximization problem subject to the energy causality constraint and the interference power constraint, and proposed the JOTPA algorithm to achieve optimal resource allocation.
By moving the multi-hop EH-CRNs around PUs, we observed that deploying the former SUs close to PUs can achieve larger end-to-end throughput and higher green energy utilization than deploying the latter SUs close to PUs. 
Moreover, by making comparisons among three different resource allocation algorithms, we concluded that JOTPA obtains larger end-to-end throughput and higher green energy utilization than ETOPA and OTEPA under all considered scenarios.

This paper provides a lower-bound for the performance of multi-hop EH-CRNs in underlay paradigm, as energy storage and management are inapplicable for the battery-free SUs. Future works considering energy storage and management will further improve the performance. Furthermore, it will be necessary to study advanced energy management schemes to cope with the dynamics of PUs if the multi-hop EH-CRNs work in interweave, overlay or even hybrid paradigm.

\begin{appendices}
\section{Proof of Proposition 1}
For the $k$-hop transmission, the achievable throughput $R_k(\tau_k, e_k)$ given in problem (\ref{equ(8)}) is the perspective of a function $\mathcal{U}(e_k) \triangleq \log_2 (1+ e_k \eta_k)$. It is obvious that $\mathcal{U}(e_k)$ is a concave function of $e_k$ since logarithmic function is concave.
As the perspective operation preserves convexity \cite{Boyd04}, $R_k(\tau_k, e_k)$ is jointly concave in $\tau_k$ and $e_k$. Furthermore, the end-to-end throughput $R(\bm{\mathcal{\tau}}, \bm{e})=\min\limits_{1 \leq k \leq K} R_k(\tau_k,e_k)$ is a jointly concave function of $\bm{\mathcal{\tau}}$ and $\bm{e}$ as $R(\bm{\mathcal{\tau}}, \bm{e})$ is the pointwise minimum of $K$ concave functions $R_k(\tau_k, e_k)$'s. The proof of Proposition 1 is thus completed.

\section{Proof of Proposition 2}
For notational convenience, we denote the throughput of each hop as $R_k$ instead of $R_k(\tau_k, e_k)$ in this proof.

First, we study the relationship between $R_k$ and $(\tau_k, e_k)$.
The partial derivative of $R_k$ with respect to $\tau_k$ is given by
\begin{equation}\label{equ(20)}
\begin{split}
\frac{\partial R_k}{\partial \tau_k}
&=\frac{1}{\ln 2}
\left( \ln \left(1+\frac{e_k \eta_k}{\tau_k}\right) -\frac{\frac{e_k \eta_k}{\tau_k}}{1+\frac{e_k \eta_k}{\tau_k}} \right)
\triangleq
\mathcal{F}(\tau_k).
\end{split}
\end{equation}
It can be verified that $\mathcal{F}(\tau_k)>0$, which means that $R_k$ monotonically increases in $\tau_k$.
Moreover, $R_k$ also monotonically increases in $e_k$. Note that $e_k$ is further constrained by the allocated time $[\tau_0,...,\tau_k]$ according to $C1$ and $C2$ in problem (\ref{equ(8)}). Thus, for given $[\tau_0,...,\tau_{k-1}]$, if $\tau_k$ is increased, $e_k$ can also be enhanced, which further improves $R_k$.
Obviously, adjusting (e.g., increasing or decreasing) $\tau_k$ should adjust $e_k$ at the same time under the constraints $C1$ and $C2$.

Then, we prove that $R_1=...=R_K$ holds for any $\bm{\mathcal{\tau}}$ satisfying $\sum_{k=0}^{K} \tau_k \leq T$.
Considering a general case that each hop obtains a different throughput, we can sort them in an ascending order as $R_1<...<R_K$ without loss of generality. Thus, the end-to-end throughput constrained by the bottleneck is $R_1$.
As $R_k$ ($k=1,...,K$) increases in $\tau_k$ and $e_k$ monotonically, by increasing ($\tau_1$, $e_1$) and decreasing ($\tau_2$, $e_2$), we can achieve $R_1'=R_2'<R_3<...<R_K$, which improves the end-to-end throughput to $R_1'$ ($R_1<R_1'<R_2$). Furthermore, by adjusting $[\tau_1,\tau_2,\tau_3]$ and $[e_1,e_2,e_3]$, we can obtain $R_1''=R_2''=R_3''<R_4<...<R_K$, which enhances the end-to-end throughput as $R_1''$ ($R_1'<R_1''<R_3$). Repeating the above process, we can improve the end-to-end throughput step by step until $R_1=...=R_K$ which cannot be further enhanced.
Thus, for the given $\bm{\mathcal{\tau}}$, we can achieve the maximum end-to-end throughput by adjusting ($\bm{\mathcal{\tau}}$,$\bm{e}$) such that each SU obtains the same throughput. 

Finally, we prove that $\sum_{k=0}^{K} \tau_k= T$ holds for the maximum end-to-end throughput $R^{*}$. Assuming $\bm{\mathcal{\tau}}'$ that satisfies $\sum_{k=0}^{K} \tau_k' < T$ and obtains $R'$, we have $\tau_K'<T-\sum_{k=0}^{K-1} \tau_k' \triangleq \tau_K''$ and further $e_K' \leq e_K''$ according to $C1$ and $C2$.
As $R_K$ is monotonically increasing in $\tau_K$ and $e_K$, there must be $R'<R''$, where $R''$ is obtained with ($\tau_K''$, $e_K''$). Then, we can always obtain a higher end-to-end throughput by adjusting $\bm{\mathcal{\tau}}$ and $\bm{e}$ as before. The above process continues until $\sum_{k=0}^{K} \tau_k= T$, which results in $R^{*}$.

\section{Proof of Proposition 3}
If $(\bm{\mathcal{\tau}}$, $\bm{e}) \in \bm{\mathcal{D}}$ is a feasible solution for problem (\ref{equ(10)}), then $R_k(\tau_k,e_k) \geq R$ holds for $k=1,...,K$. As $\bm{\mathcal{\lambda}} \geq 0$, it is obvious that $\mathcal{L}(\bm{\mathcal{\tau}},\bm{e},\bm{\mathcal{\lambda}}) \leq 0$ by (\ref{equ(11)}), and $\mathcal{G}(\bm{\mathcal{\lambda}}) \leq 0$ as $\mathcal{G}(\bm{\mathcal{\lambda}})$ is the minimum of all $\mathcal{L}(\bm{\mathcal{\tau}},\bm{e},\bm{\mathcal{\lambda}})$ by (\ref{equ(12)}). This is consistent with the ``if" part of Proposition 3.

Then, we prove the ``only if" part by contradiction. We assume that $(\bm{\mathcal{\tau}}$, $\bm{e}) \in \bm{\mathcal{D}}$ is a feasible solution for problem (\ref{equ(10)}) and there exists $\bm{\mathcal{\lambda}} \geq 0$ satisfying $\mathcal{G}(\bm{\mathcal{\lambda}}) > 0$. Thus, $\lambda_k (R_k(\tau_k,e_k)-R) \geq 0$ and $\mathcal{G}(\bm{\mathcal{\lambda}}) \leq \mathcal{L}(\bm{\mathcal{\tau}},\bm{e},\bm{\mathcal{\lambda}}) \leq 0$, which conflicts with the assumption that $\mathcal{G}(\bm{\mathcal{\lambda}}) > 0$. This completes the ``only if" part of Proposition 3.

\section{Proof of Proposition 4}

As problem (\ref{equ(13)}) is a convex optimization problem, there is strong duality between the primal problem and the dual problem by Slater's condition.
Therefore, according to the Karush-Kuhn-Tucker (KKT) condition \cite{Boyd04}, the optimal solution must satisfy $\frac{\partial \mathcal{L}'(\bm{\mathcal{\tau}}, \bm{e}, \bm{\mathcal{\mu}}, \mathcal{\omega})}{\partial \tau_k} | _{\tau_k=\tau_k^*}=0$ $(k=0,...,K)$ for given $\bm{e}$,
and $\frac{\partial \mathcal{L}'(\bm{\mathcal{\tau}}, \bm{e}, \bm{\mathcal{\mu}}, \mathcal{\omega})}{\partial e_k}| _{e_k=e_k^*}=0$ $(k=1,...,K)$ for given $\bm{\mathcal{\tau}}$.

On one hand, given $\bm{e}$, by calculating $\frac{\partial \mathcal{L}'(\bm{\mathcal{\tau}}, \bm{e}, \bm{\mathcal{\mu}}, \mathcal{\omega})}{\partial \tau_k}=0$ for $k=0,...,K$, we have
\begin{equation}\label{equ(21)}
\begin{split}
\left\{
               \begin{array}{l}
               \xi P_t \sum\limits_{j=1}^{K} \mu_j g_{E,j} - \omega=0, \quad k=0,
               \\
               \lambda_k \mathcal{F}(\tau_k) + \xi P_t \sum\limits_{j=k+1}^{K} \mu_j g_{E,j} -\omega=0,
               \quad k=1,...,K-1,
               \\
               \lambda_k  \mathcal{F}(\tau_k)  -\omega=0, \quad k=K.
               \end{array}
             \right.
\end{split}
\end{equation}

Combining the equations in (\ref{equ(21)}), we can omit $\mathcal{\omega}$ and obtain
$\lambda_k \mathcal{F}(\tau_k)=\xi P_t \sum\limits_{j=1}^{k} \mu_j g_{E,j}$, i.e.,
\begin{equation}\label{equ(22)}
\begin{split}
\ln \left(1+\frac{e_k \eta_k}{\tau_k}\right) -\frac{\frac{e_k \eta_k}{\tau_k}}{1+\frac{e_k \eta_k}{\tau_k}}=\frac{\ln 2}{\lambda_k} \xi P_t \sum\limits_{j=1}^{k} \mu_j g_{E,j }.
\end{split}
\end{equation}

After some mathematical manipulations, we employ the Lambert W function which is the inverse function of $f(t)=t \exp (t)$ \cite{Corless96} to solve $\tau_k$ for $k=1,...,K$. The obtained solution is given by (\ref{equ(15)}), with which we can calculate $\tau_0$ as (\ref{equ(16)}) by Proposition 2.

On the other hand, given $\bm{\mathcal{\tau}}$, by calculating $\frac{\partial \mathcal{L}'(\bm{\mathcal{\tau}}, \bm{e}, \bm{\mathcal{\mu}}, \mathcal{\omega})}{\partial e_k}=0$ for $k=1,...,K$, we have
\begin{equation}\label{equ(23)}
\begin{split}
e_k=\frac{\tau_k}{\eta_k}\left(\frac{\lambda_k \eta_k}{\ln 2 \mu_k}-1\right).
\end{split}
\end{equation}
As the allocated energy $e_k$ is also subject to the interference power constraint (i.e., $0 \leq e_k \leq \frac{I_p \tau_k}{g_{I,k}}$), we have the optimal energy allocation as (\ref{equ(17)}).

\end{appendices}




\end{document}